\documentclass[journal,twoside,web]{ieeecolor}
\usepackage{generic}
\usepackage{cite}
\usepackage{graphicx}
\usepackage{algorithm}
\usepackage{algorithmic}
\usepackage{amsmath,amssymb,amsfonts,mathtools}
\usepackage{hyperref}
\hypersetup{hidelinks=true}
\usepackage{textcomp}
\usepackage{multirow}
\usepackage{booktabs}
\usepackage{xspace}
\makeatletter
\DeclareRobustCommand\onedot{\futurelet\@let@token\@onedot}
\def\@onedot{\ifx\@let@token.\else.\null\fi\xspace}

\def\eg{\emph{e.g}\onedot}

\def\etal{\emph{et al}\onedot}
\makeatother
\usepackage[capitalize]{cleveref}
\def\BibTeX{{\rm B\kern-.05em{\sc i\kern-.025em b}\kern-.08em
    T\kern-.1667em\lower.7ex\hbox{E}\kern-.125emX}}
\begin{document}
\title{
High-Fidelity Functional\\Ultrasound Reconstruction via\\A Visual Auto-Regressive Framework
}

\author{Xuhang Chen, Zhuo Li, Yanyan Shen, Mufti Mahmud, Hieu Pham, Chi-Man Pun and Shuqiang Wang
\thanks{This work was supported by National Natural Science Foundations of China under Grant 62172403, 12326614.}
\thanks{Xuhang Chen, Zhuo Li, Yanyan Shen and Shuqiang Wang are with the Shenzhen Institutes of Advanced Technology, Chinese Academy of Sciences, Shenzhen 518055, China (e-mail: xx.chen2@siat.ac.cn; z.li6@siat.ac.cn; yy.shen@siat.ac.cn; sq.wang@siat.ac.cn).}
\thanks{Mufti Mahmud is with Department of Information and Computer Science, SDAIA-KFUPM Joint Research Center for AI, Interdisciplinary Research Center for Biosystems and Machines, King Fahd University of Petroleum and Minerals, Dhahran, Saudi Arabia (email: mufti.mahmud@kfupm.edu.sa).}
\thanks{Xuhang Chen and Chi-Man Pun are with the Department of Computer and Information Science, University of Macau, Macau SAR 999078, China (email: yc17491@umac.mo; cmpun@umac.mo).}
\thanks{Hieu Pham is with the College of Engineering and Computer Science and the VinUni-Illinois Smart Health Center, VinUniversity, Hanoi 100000, Vietnam (email: hieu.ph@vinuni.edu.vn).}
\thanks{Chi-Man Pun and Shuqiang Wang are the corresponding authors (e-mail: cmpun@umac.mo; sq.wang@siat.ac.cn).}
}

\maketitle

\begin{abstract}
Functional ultrasound (fUS) imaging provides exceptional spatiotemporal resolution for neurovascular mapping, yet its practical application is significantly hampered by critical challenges. Foremost among these are data scarcity, arising from ethical considerations and signal degradation through the cranium, which collectively limit dataset diversity and compromise the fairness of downstream machine learning models. To address these limitations, we introduce UltraVAR (Ultrasound Visual Auto-Regressive model), the first data augmentation framework designed for fUS imaging that leverages a pre-trained visual auto-regressive generative model. UltraVAR is designed not only to mitigate data scarcity but also to enhance model fairness through the reconstruction of diverse and physiologically plausible fUS samples. The generated samples preserve essential neurovascular coupling features—specifically, the dynamic interplay between neural activity and microvascular hemodynamics. This capability distinguishes UltraVAR from conventional augmentation techniques, which often disrupt these vital physiological correlations and consequently fail to improve, or even degrade, downstream task performance. The proposed UltraVAR employs a scale-by-scale reconstruction mechanism that meticulously preserves the spatial topological relationships within vascular networks. The framework's fidelity is further enhanced by two integrated modules: the Smooth Scaling Layer, which ensures the preservation of critical image information during multi-scale feature propagation, and the Perception Enhancement Module, which actively suppresses artifact generation via a dynamic residual compensation mechanism. Comprehensive experimental validation demonstrates that datasets augmented with UltraVAR yield statistically significant improvements in downstream classification accuracy. This work establishes a robust foundation for advancing ultrasound-based neuromodulation techniques and brain-computer interface technologies by enabling the reconstruction of high-fidelity, diverse fUS data.
\end{abstract}


\section{Introduction}
\IEEEPARstart{F}{unctional} ultrasound imaging (fUS) has emerged as a transformative neuroimaging in neuroscience, offering exceptional spatiotemporal resolution capabilities—with spatial resolution at the micrometer scale and temporal resolution reaching millisecond levels \cite{10314736,10388392}. However, fUS imaging faces significant  challenges that hinder its widespread clinical adoption. First, the substantial attenuation of ultrasound waves by cranial structures severely degrades signal quality in non-invasive applications. Second, the acquisition of human brain fUS data encounters formidable obstacles due to ethical constraints and procedural complexities associated with invasive techniques, resulting in limited dataset sizes and inadequate demographic representation \cite{rabut2024functional,Brunner2021WholebrainFU,DiIanni2022DeepfUSAD}. These limitations constrain fUS's potential for advancing neurophysiological research \cite{10663841,renaudin2022functional,sieu2015eeg} and clinical translation \cite{8476975,10374553,Norman2020SingletrialDO,zhu2021integrating}.

Recent advances in generative artificial intelligence (Gen-AI) have demonstrated impressive effectiveness across various medical imaging tasks, including disease diagnosis \cite{zuo2024prior,10599152}, super-resolution \cite{you2022fine,10839074,8859355,you2022brain}, cross-modal translation \cite{li2025scdm,yao2025catd,hu2021bidirectional}, brain network construction \cite{zong2024new,jing2024estimating,10562194,kong2022adversarial,gong2023addictive,jing2024addiction} and image synthesis \cite{wang2024enhanced}. To address fUS-specific challenges, UltraVAR (Ultrasound Visual Auto-Regressive model) is proposed as the first data augmentation framework for fUS imaging based on a pre-trained visual auto-regressive generative model (VAR) \cite{tian2024visual}. The proposed method addresses not only the data scarcity but also enhances model fairness by generating realistic samples \cite{zhang2025towards,hu2024enhancing,xing2025achieving}. The proposed UltraVAR model demonstrates effectiveness in preserving the essential neurovascular coupling features that underpin fUS technology—the dynamic relationship between neural activity and microvascular blood volume fluctuations. While conventional data augmentation techniques (\eg, geometric transformations or noise injection) frequently disrupt these critical physiological correlations and reduce model fairness, the pre-trained VAR model employs a scale-by-scale prediction mechanism that accurately preserves the spatial topological relationships of vascular networks across cerebral regions. The model maintains visual patterns between critical structures such as the primary motor cortex and posterior parietal cortex, ensuring generated samples retain both hemodynamic characteristics and reflect neurophysiological diversity across populations. This capability proves particularly valuable for emerging research domains such as neonatal brain development monitoring, yielding superior fUS image quality and enhanced blood flow signal reconstruction, thereby providing a more robust and representative imaging foundation for applications in brain function decoding and brain-computer interfaces (BCIs).

To further enhance UltraVAR's performance, two new components are proposed: the Smooth Scaling Layer and the Perception Enhancement Module. The Smooth Scaling Layer preserves critical image information during multi-scale feature propagation, while the Perception Enhancement Module effectively suppresses artifact generation through a dynamic residual compensation mechanism. This dual optimization produces fUS images characterized by exceptional visual appearance and anatomical fidelity. These improvements not only establish a high-quality data foundation for ultrasound neuromodulation techniques but also accelerate the clinical translation of ultrasound-based brain-computer interface technologies. Our experimental validation demonstrates that datasets augmented with UltraVAR yield significant improvements in downstream classification task accuracy compared to alternative methods, underscoring the transformative potential of our approach in advancing clinical applications of fUS technology.

The key contributions of this work can be summarized as follows:
\begin{enumerate}
    \item UltraVAR, a visual auto-regressive framework is proposed for functional ultrasound image augmentation that preserves neurovascular coupling features.
    \item Smooth Scaling Layer is proposed to maintain information integrity during multi-scale feature propagation, ensuring coherent transitions between different patch resolutions in the hierarchical reconstruction process.
    \item Perception Enhancement Module is proposed to significantly reduce artifacts and enhance visual quality in the augmented fUS images.
\end{enumerate}

\section{Related Work}
\subsection{Functional Ultrasound Imaging}
Functional ultrasound imaging has rapidly emerged as a pivotal neuroimaging technique, distinguished by its exceptional spatiotemporal resolution in monitoring cerebrovascular dynamics and its adaptability to diverse experimental conditions, including awake, head-fixed, or freely behaving animal models \cite{montaldo2022functional}. The foundational study by Mac{'e} \etal \cite{mace2011functional} first established fUS as a potent tool for imaging brain function, demonstrating its capacity to map cerebral blood flow changes intrinsically linked to neural activity in small-animal models, thereby validating its sensitivity to neurovascular coupling mechanisms.

Building upon this preclinical success, fUS has seen remarkable translation into human applications. These include non-invasive bedside monitoring of brain activity and dynamic cortical connectivity in neonates \cite{demene2017functional, baranger2021bedside}, offering critical insights into early neurodevelopment and enabling continuous assessment of brain states. Furthermore, its utility extends to complex clinical and research scenarios, such as providing intraoperative guidance during awake craniotomies by mapping functional and vascular territories \cite{soloukey2020functional}, decoding functional states of spinal cord networks to inform the development of real-time closed-loop neuromodulation systems \cite{agyeman2024functional}, and facilitating the emergence of non-invasive real-time brain-computer interfaces \cite{zheng2023emergence}. These diverse applications underscore the expanding versatility of fUS across both fundamental neuroscience research and clinical practice.

Recent innovations have specifically targeted key translational challenges, particularly those related to imaging the human brain in more ecologically valid settings. Addressing the limitations imposed by the skull and motion artifacts, Rabut \etal \cite{rabut2024functional} introduced an innovative acoustically transparent cranial window. This advancement enables high-resolution fUS imaging in freely moving human subjects, significantly mitigating motion-induced distortions. Complementing such efforts to improve imaging through the skull, Soloukey \etal \cite{soloukey2024p09} have explored skull-implant-compatible fUS configurations, thereby extending the potential for chronic neuroimaging in unconstrained human environments. Collectively, these pioneering developments chart the evolution of fUS from a specialized preclinical modality to an increasingly versatile and accessible platform for investigating brain function under naturalistic conditions, promising new avenues for both fundamental discovery and clinical intervention.

\subsection{Visual Auto-Regressive Models}
In parallel, visual auto-regressive modeling has revolutionized data augmentation by enabling scalable, high-fidelity image synthesis. Tian \etal \cite{tian2024visual} pioneered a next-scale prediction framework that iteratively refines image details, achieving superior performance in resolution and semantic coherence. Complementary approaches, such as STAR \cite{ma2024star} and M-VAR \cite{ren2024m}, leverage hierarchical auto-regressive representations: STAR integrates text conditioning to guide scale-wise generation, while M-VAR decouples spatial and contextual dependencies to enhance detail preservation. Concurrently, tokenization frameworks like Imagefolder \cite{li2024imagefolder} and XQ-GAN \cite{li2024xq} optimize latent space discretization, streamlining training and inference for large-scale datasets.

These advancement in visual auto-regressive modeling not only push the boundaries of image reconstruction but also provide valuable insights into the integration of generative models with fUS. Recent work has also explored diffusion-based approaches for medical data augmentation \cite{yao2023conditional}, highlighting the growing importance of generative models in addressing data scarcity challenges. By combining these domains, future research can explore novel approaches to augmenting and interpreting functional ultrasound data, paving the way for more robust and interpretable neurophysiological insights.

\section{Methods}
\begin{figure*}[ht]
    \centering
    \includegraphics[width=\linewidth]{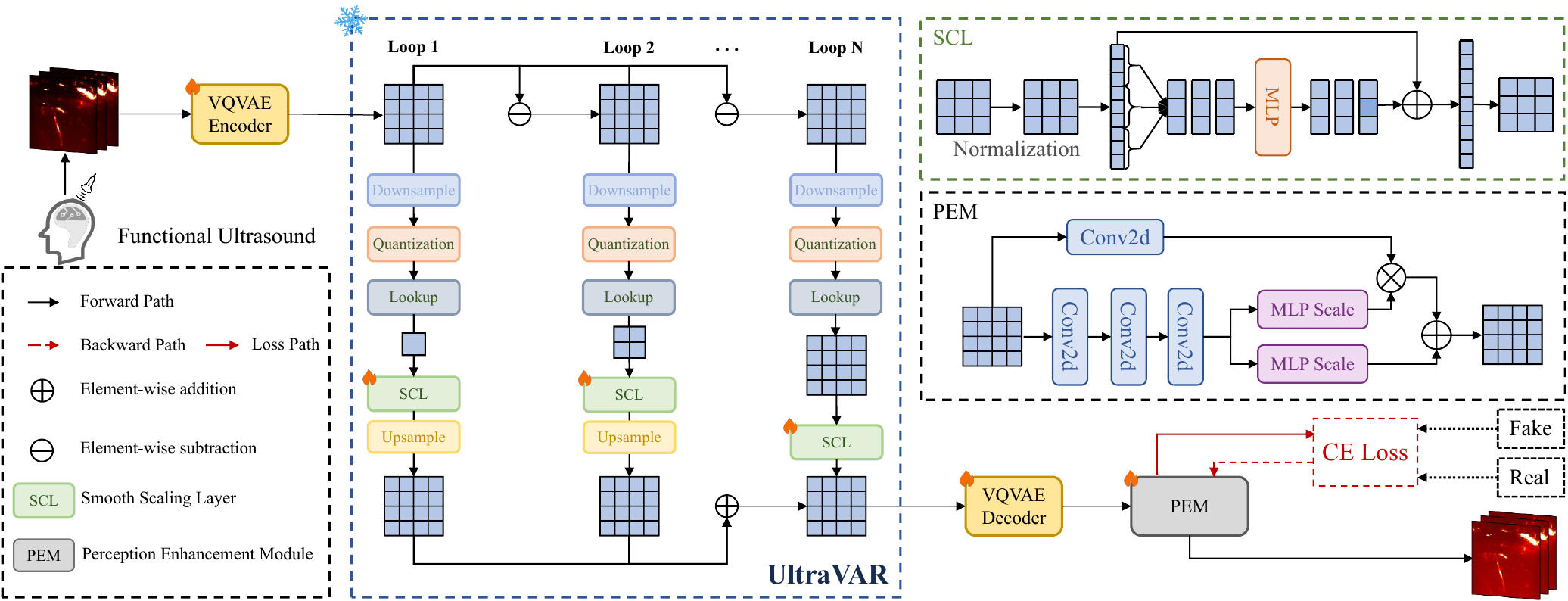}
    \caption{Overview of the proposed UltraVAR framework. The architecture consists of a VQVAE backbone with encoder and decoder components, integrated with two  modules: the Smooth Scaling Layer (SCL) and Perception Enhancement Module (PEM). The hierarchical reconstruction process refines image patches at progressive resolutions.}
    \label{fig:model}
\end{figure*}

\subsection{Overview}
The proposed UltraVAR framework addresses the challenge of functional ultrasound image augmentation through a hierarchical generative modeling approach as shown in \cref{fig:model}. The proposed method leverages a Vector Quantized Variational Autoencoder (VQVAE) backbone \cite{Oord2017NeuralDR}  with a Vision Auto-Regressive model for sequential reconstruction of image patches at progressively higher resolutions. To enhance the quality and fidelity of the generated functional ultrasound images, we propose two new modules: (1) Smooth Scaling Layer that ensures coherent transitions between different patch resolutions, and (2) Perception Enhancement Module that improves the visual characteristics of the decoded images. The overall architecture produces high-quality functional ultrasound images with controlled variations based on conditional inputs.

\subsection{VQVAE Backbone and Pre-trained VAR Model}
Our framework utilizes a VQVAE architecture as its backbone for learning a compressed latent representation of ultrasound images. The VQVAE consists of an encoder, a vector quantizer, and a decoder. The encoder transforms an input image $x \in \mathbb{R}^{1 \times H \times W}$ into a latent representation $f \in \mathbb{R}^{C \times h \times w}$, where $C$ is the number of channels in the latent space, and $h$, $w$ are the spatial dimensions of the latent representation.

The vector quantizer discretizes this continuous latent representation by mapping each spatial vector to its nearest neighbor in a learned codebook $\mathcal{Z} = \{z_k\}_{k=1}^K$, where $K$ is the vocabulary size. This quantization process produces a sequence of discrete indices that can be modeled auto-regressively.

The VAR component operates on these quantized representations with a hierarchical approach using multiple patch sizes $\{p_1, p_2, ..., p_n\}$ (ranging from 1 to 32 in our implementation). This multi-scale approach enables the model to capture both global structure and fine local details of the ultrasound images. The auto-regressive reconstruction proceeds by predicting patches of increasing resolution, with each stage conditioning on previously generated lower-resolution patches. 

The proposed model uses a sequence of self-attention layers \cite{vaswani2017attention} with rotary positional embeddings to process the token representations. Conditional reconstruction is achieved by incorporating class embeddings that guide the reconstruction process. We employ classifier-free guidance during inference to strengthen the conditioning signal, with the guidance scale adjusted proportionally to the reconstruction stage.

\begin{algorithm*}[ht]
\caption{Pseudocode for Training and Inference with UltraVAR}
\label{alg:ultravar}
\begin{algorithmic}[1]
\REQUIRE Training dataset $\mathcal{D}=\{(x_i, c_i)\}$, VQVAE parameters $\theta_{\mathrm{vq}}$, UltraVAR parameters $\theta_{\mathrm{var}}$, number of epochs $N_{\mathrm{epochs}}$.
\STATE \textbf{Stage 1: VQVAE Training}
\STATE Initialize VQVAE with configuration $(\mathrm{Encoder}, \mathrm{Decoder}, \mathrm{Quantizer})$.
\FOR{epoch $=1$ to $N_{\mathrm{epochs}}$}
    \FOR{each mini-batch $(x,c)$ in $\mathcal{D}$}
        \STATE $(\hat{x}, \mathbf{r}, \mathrm{idxs}, \mathbf{scales}, \mathcal{L}_{\mathrm{quant}}) \leftarrow \mathrm{VQVAE}(x)$
        \STATE $\mathcal{L}_{\mathrm{recon}} \leftarrow \|\hat{x} - x\|^2 \quad\;\;$ \COMMENT{MSE reconstruction}
        \STATE $\mathcal{L}_{\mathrm{VQVAE}} \leftarrow \mathcal{L}_{\mathrm{rec}} + \mathcal{L}_{\mathrm{quant}}$
        \STATE Update $\theta_{\mathrm{vq}}$ by backprop with $\mathcal{L}_{\mathrm{total}}$
    \ENDFOR
\ENDFOR
\STATE Freeze VQVAE parameters $\theta_{\mathrm{vq}}^\ast \leftarrow \theta_{\mathrm{vq}}$.

\STATE \textbf{Stage 2: UltraVAR Fine-tuning}
\STATE Initialize UltraVAR with $(\mathrm{VQVAE}(\theta_{\mathrm{vq}}^\ast), \mathrm{TransformerLayers}, \mathrm{FinalLayer}, \mathrm{SmoothScalingLayer})$.
\FOR{epoch $=1$ to $N_{\mathrm{epochs}}$}
    \FOR{each mini-batch $(x,c)$ in $\mathcal{D}$}
        \STATE \COMMENT{Step A: Encode image to latent indices}
        \STATE $(\cdot,\cdot,\mathrm{idxs}_R,\mathbf{scales}_{BLC},\cdot) \leftarrow \mathrm{VQVAE}(x)$
        \STATE $\mathbf{x}_{\mathrm{AR}} \leftarrow \mathrm{Concat}(\mathrm{idxs}_R)$ \COMMENT{Flatten or reshape latent tokens}
        \STATE \COMMENT{Step B: Auto-regressive prediction via UltraVAR}
        \STATE $\mathbf{\hat{z}}_{\mathrm{logits}} \leftarrow \mathrm{UltraVAR}\bigl(\mathbf{x}_{\mathrm{AR}}, c\bigr)$ 
        \STATE $\mathcal{L}_{\mathrm{VAR}} \leftarrow \mathrm{CrossEntropy}\bigl(\mathbf{\hat{z}}_{\mathrm{logits}}, \mathbf{x}_{\mathrm{AR}}\bigr)$
        \STATE \COMMENT{Step C: Backprop and optimize UltraVAR}
        \STATE Update $\theta_{\mathrm{var}}$ by backprop with $\mathcal{L}_{\mathrm{VAR}}$
    \ENDFOR
\ENDFOR

\STATE \textbf{Stage 3: Inference}
\STATE \COMMENT{Given a condition $c$, generate new output}
\STATE \textbf{function} $\mathrm{UltraVAR}.generate(c, \mathrm{cfg\_scale}, \mathrm{temp}, \mathrm{top\_p})$:
\STATE \quad Construct empty latent feature map $\mathbf{f}_{\mathrm{out}} \leftarrow \mathbf{0}$
\STATE \quad \textbf{for} patch stage $i$ from $1$ to $|\mathrm{patch\_sizes}|$ \textbf{do}
\STATE \qquad \COMMENT{Compute stage ratio for classifier-free guidance}
\STATE \qquad $\mathrm{stage\_ratio} = i / (\mathrm{num\_stages} - 1)$
\STATE \qquad $\mathbf{\hat{z}}_{\mathrm{logits}} \leftarrow \mathrm{PredictNextTokens}(\mathbf{f}_{\mathrm{out}}, c, \mathrm{stage\_ratio})$
\STATE \qquad $\mathbf{\hat{z}}_{\mathrm{idx}} \leftarrow \mathrm{Sample}\bigl(\mathbf{\hat{z}}_{\mathrm{logits}}, \mathrm{temp}, \mathrm{top\_p}\bigr)$
\STATE \qquad \COMMENT{Lookup code embeddings, interpolate to final size}
\STATE \qquad $\mathbf{z}_q \leftarrow \mathrm{VQVAE.codebook}(\mathbf{\hat{z}}_{\mathrm{idx}})$
\STATE \qquad $\mathbf{z}_q \leftarrow \mathrm{Interpolate}(\mathbf{z}_q, \mathrm{final\_patch\_size})$
\STATE \qquad $\mathbf{f}_{\mathrm{out}} \leftarrow \mathbf{f}_{\mathrm{out}} + \mathrm{phi\_update}(\mathbf{z}_q, \mathrm{stage\_ratio})$
\STATE \quad \textbf{end for}
\STATE \quad \textbf{return} $\mathbf{f}_{\mathrm{out}}$

\end{algorithmic}
\end{algorithm*}

\subsection{Smooth Scaling Layer}
To enhance the coherence between patches of different resolutions and maintain spatial consistency across generated images, we introduce the Smooth Scaling Layer. This component plays a crucial role in ensuring smooth transitions between different patch resolutions during the hierarchical reconstruction process.

The Smooth Scaling Layer processes the logits produced by the transformer model before they are used for token sampling. It operates on windows of fixed size (set to 8 in our implementation), applying a learned transformation that promotes local coherence:
\begin{equation}
\text{SSL}(x) = x + \text{MLP}(\text{LayerNorm}(x)),
\end{equation}
where $x \in \mathbb{R}^{B \times L \times V}$ represents the logits for a batch of $B$ sequences with length $L$ and vocabulary size $V$. The MLP consists of a two-layer feedforward network with GELU activation and dropout for regularization:
\begin{equation}
\text{MLP}(x) = W_2(\text{GELU}(W_1 x)).
\end{equation}

By processing the logits in fixed-size windows, the Smooth Scaling Layer enforces local consistency in the predicted token distributions, which is especially important for maintaining structural coherence in medical ultrasound images. The layer adaptively refines the predicted distributions based on neighboring logits, resulting in more realistic and visually coherent image reconstruction.

\subsection{Perception Enhancement Module}
To further improve the visual quality of the generated ultrasound images, we introduce the Perception Enhancement Module (PEM). While the VQVAE decoder reconstructs images from the quantized latent representations, PEM acts as a refinement module that enhances the perceptual quality of the final outputs, similar to enhancement approaches in other domains \cite{chen2024uwformer}.

PEM adopts a feature modulation approach through Global Feature Modulation (GFM) modules, which adaptively adjust features based on image-specific conditions. The network first extracts a condition vector from the input image through a ConditionNet:
\begin{equation}
\text{cond} = \text{ConditionNet}(x).
\end{equation}

This condition vector is then used to modulate features through a series of GFM modules:
\begin{equation}
\text{GFM}(x, \text{cond}) = \text{ReLU}(f(x) \cdot \text{scale}(\text{cond}) + \text{shift}(\text{cond}) + f(x)),
\end{equation}
where $f(x)$ is a convolutional projection of the input, and scale and shift are learned transformations of the condition vector.

The PEM consists of three cascaded GFM modules, which progressively enhance the visual characteristics of the images. This modulation-based approach enables the network to learn image-specific enhancements that improve contrast, sharpness, and overall visual fidelity in the generated ultrasound images, drawing inspiration from recent advances in image restoration \cite{guo2025underwater}. By incorporating this perception-aware module, our framework generates not only structurally accurate but also visually enhanced ultrasound images that better highlight the relevant anatomical features.

\subsection{Objective Function}
The training of UltraVAR proceeds in two stages. First, we train the VQVAE backbone with a combination of reconstruction and quantization losses:
\begin{equation}
\mathcal{L}_{\text{VQVAE}} = \mathcal{L}_{\text{recon}} + \mathcal{L}_{\text{quant}},
\end{equation}
where $\mathcal{L}_{\text{recon}} = \|x - \hat{x}\|_2^2$ is the mean squared error between the input and reconstructed images, and $\mathcal{L}_{\text{quant}}$ is the quantization loss that includes both codebook loss and commitment loss:
\begin{equation}
\mathcal{L}_{\text{quant}} = \|\text{sg}[\hat{f}] - f\|_2^2 + \beta\|\hat{f} - \text{sg}[f]\|_2^2,
\end{equation}
where $\text{sg}[\cdot]$ is the stop-gradient operator, $f$ is the encoder output, $\hat{f}$ is the quantized representation, and $\beta$ is a weighting factor set to 0.25.

Once the VQVAE is trained, we freeze its parameters and train the VAR model using a cross-entropy loss for predicting the next token in the sequence:
\begin{equation}
\mathcal{L}_{\text{VAR}} = \text{CrossEntropy}(\text{logits}, \text{idx}),
\end{equation}
where logits are the model's predictions and idx are the indices from the VQVAE encoding. This auto-regressive training allows the model to learn the distribution of token sequences that correspond to valid ultrasound images.

\section{Experiment}

\begin{figure*}
    \centering
    \includegraphics[width=\linewidth]{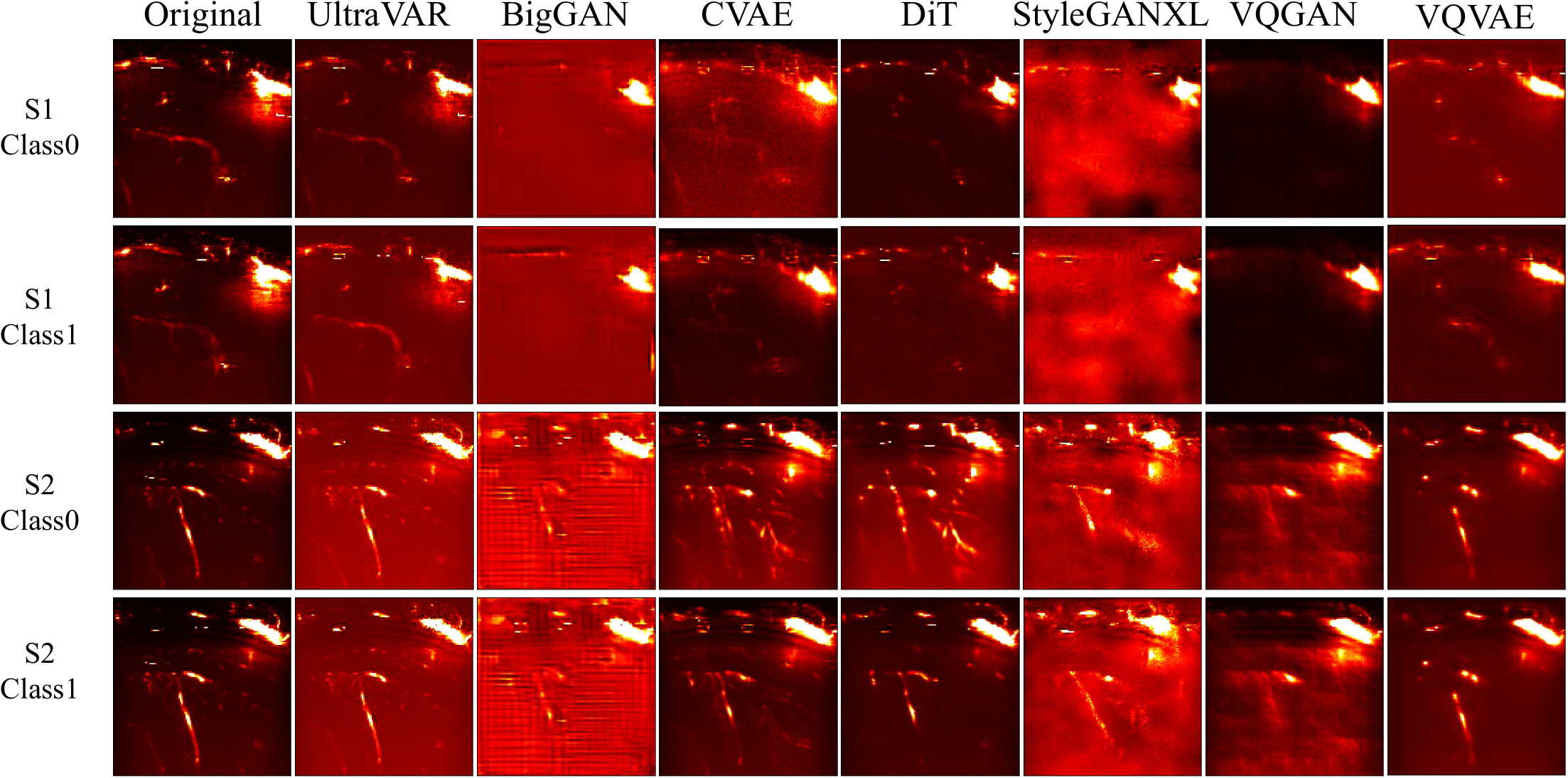}
    \caption{Visual comparison of fUS images. This figure displays original fUS images alongside synthetic samples generated by UltraVAR and various baseline methods. Comparisons are shown for different brain activity classes (Class0 and Class1) across two experimental sessions (S1 stands for Session1 and S2 stands for Session2).}
    \label{fig:vis}
\end{figure*}

\begin{figure*}
    \centering
    \includegraphics[width=0.32\linewidth]{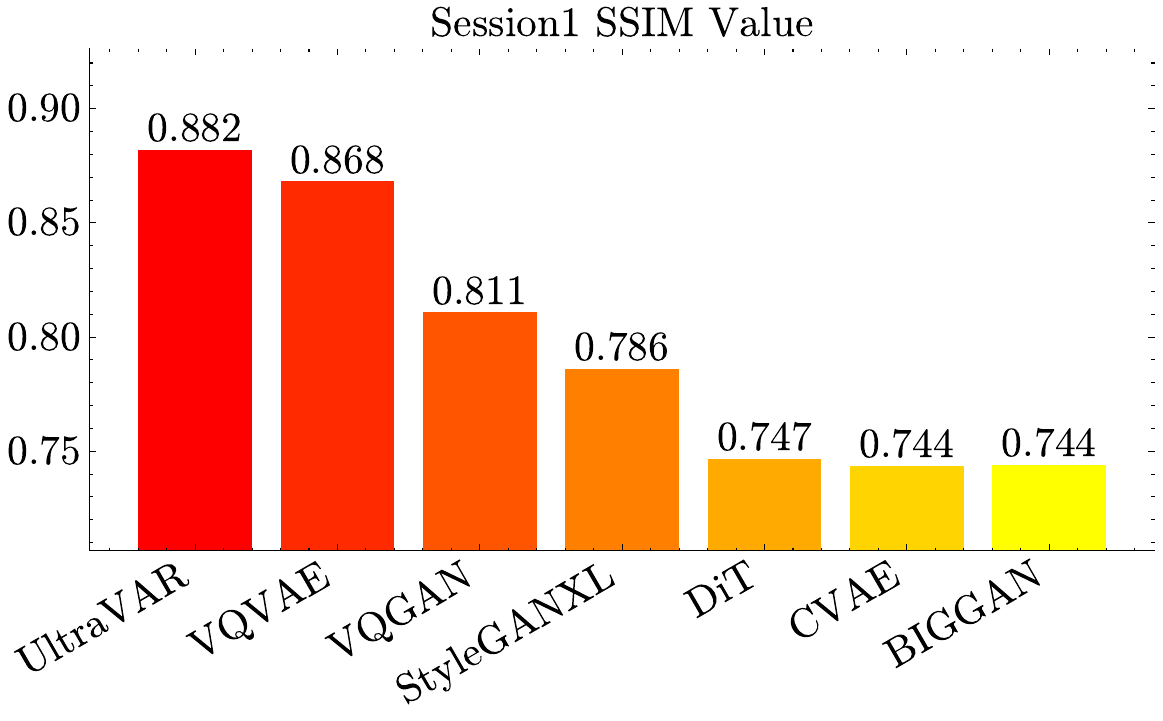}
    \includegraphics[width=0.32\linewidth]{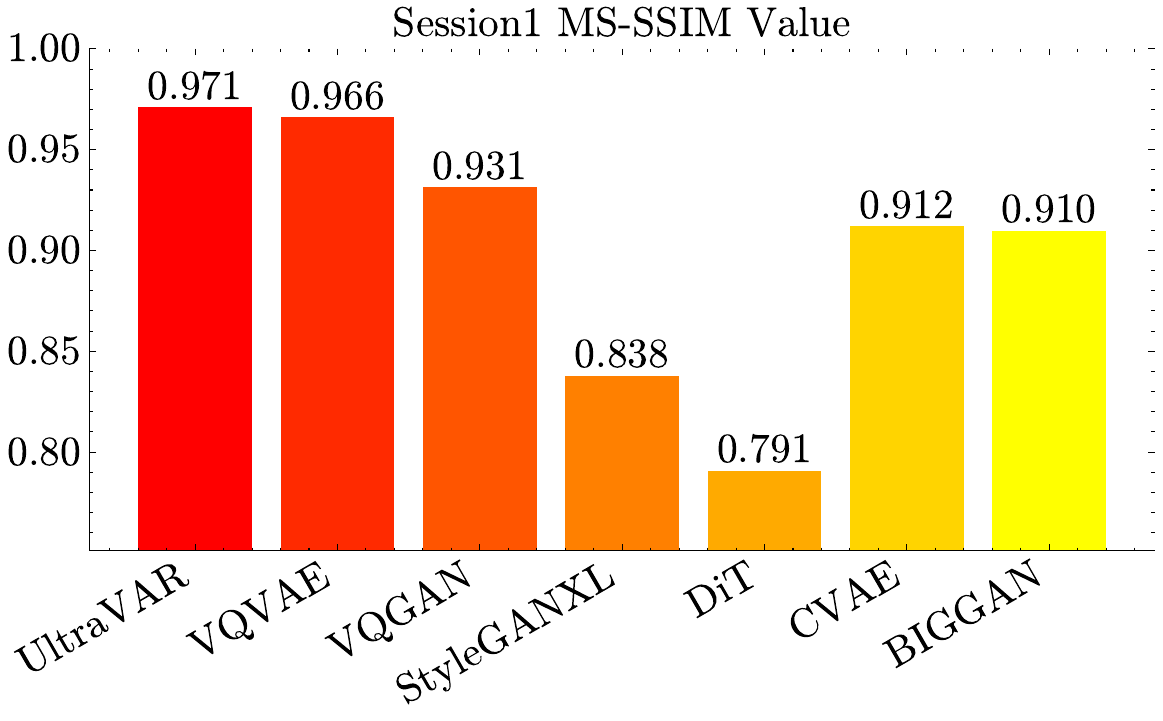}
    \includegraphics[width=0.32\linewidth]{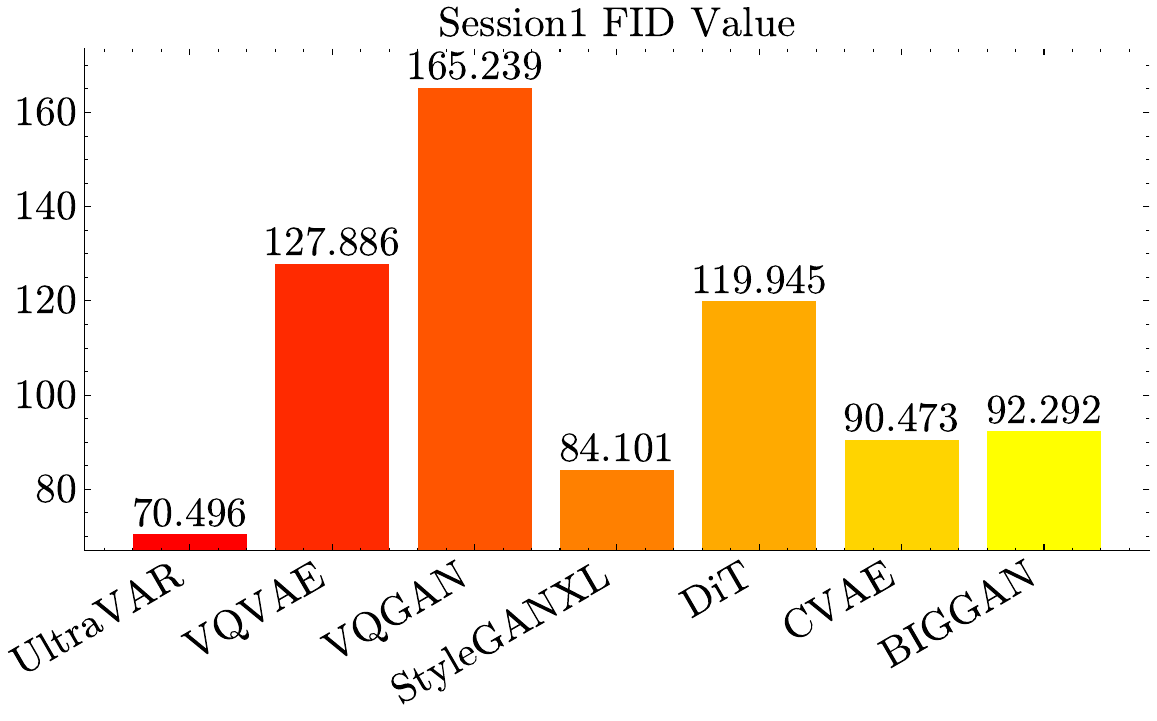}
    \includegraphics[width=0.32\linewidth]{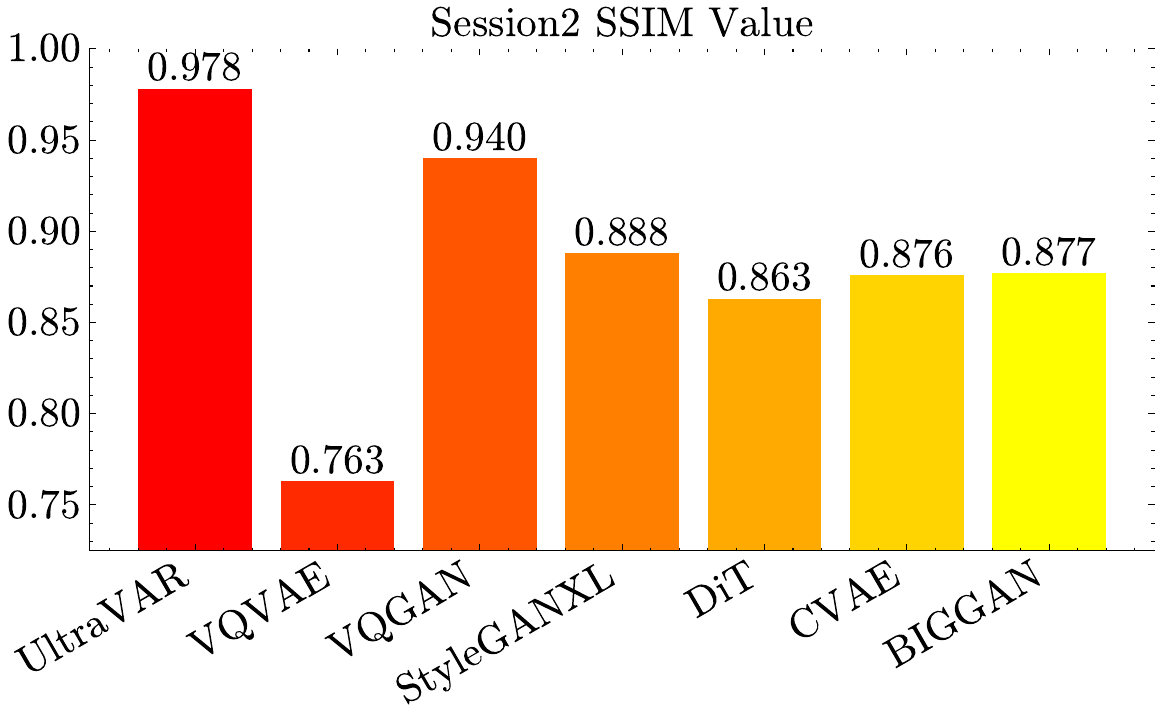}
    \includegraphics[width=0.32\linewidth]{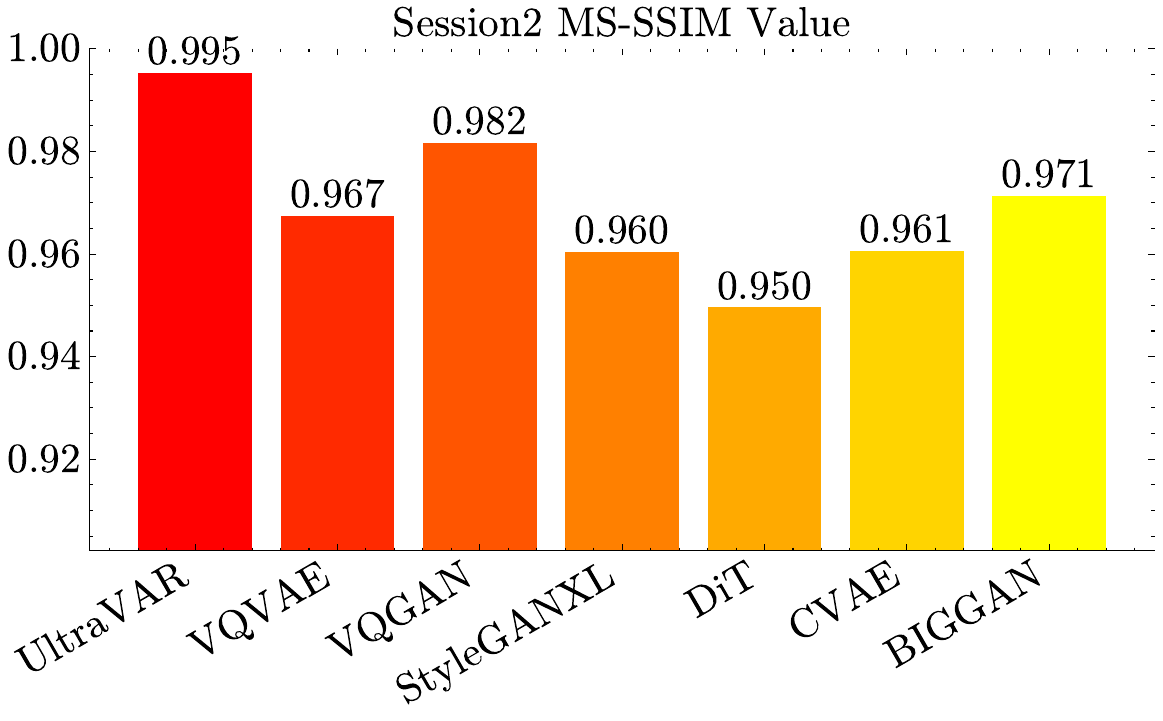}
    \includegraphics[width=0.32\linewidth]{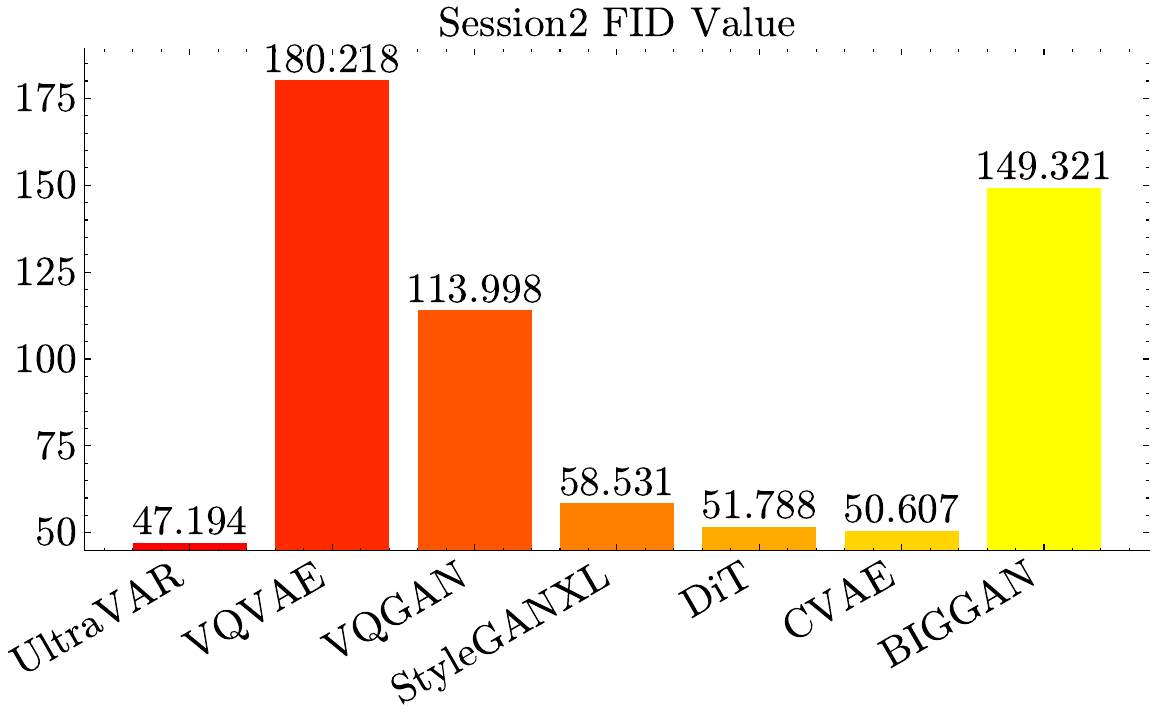}
    \caption{Quantitative comparison of image quality metrics across different generative models. The charts display SSIM, MS-SSIM, and FID values for Session1 (top row) and Session2 (bottom row). UltraVAR consistently outperforms other methods with higher SSIM and MS-SSIM values and lower FID scores.
    }
    \label{fig:comp}
\end{figure*}

\begin{figure*}
    \centering
    \includegraphics[width=0.48\linewidth]{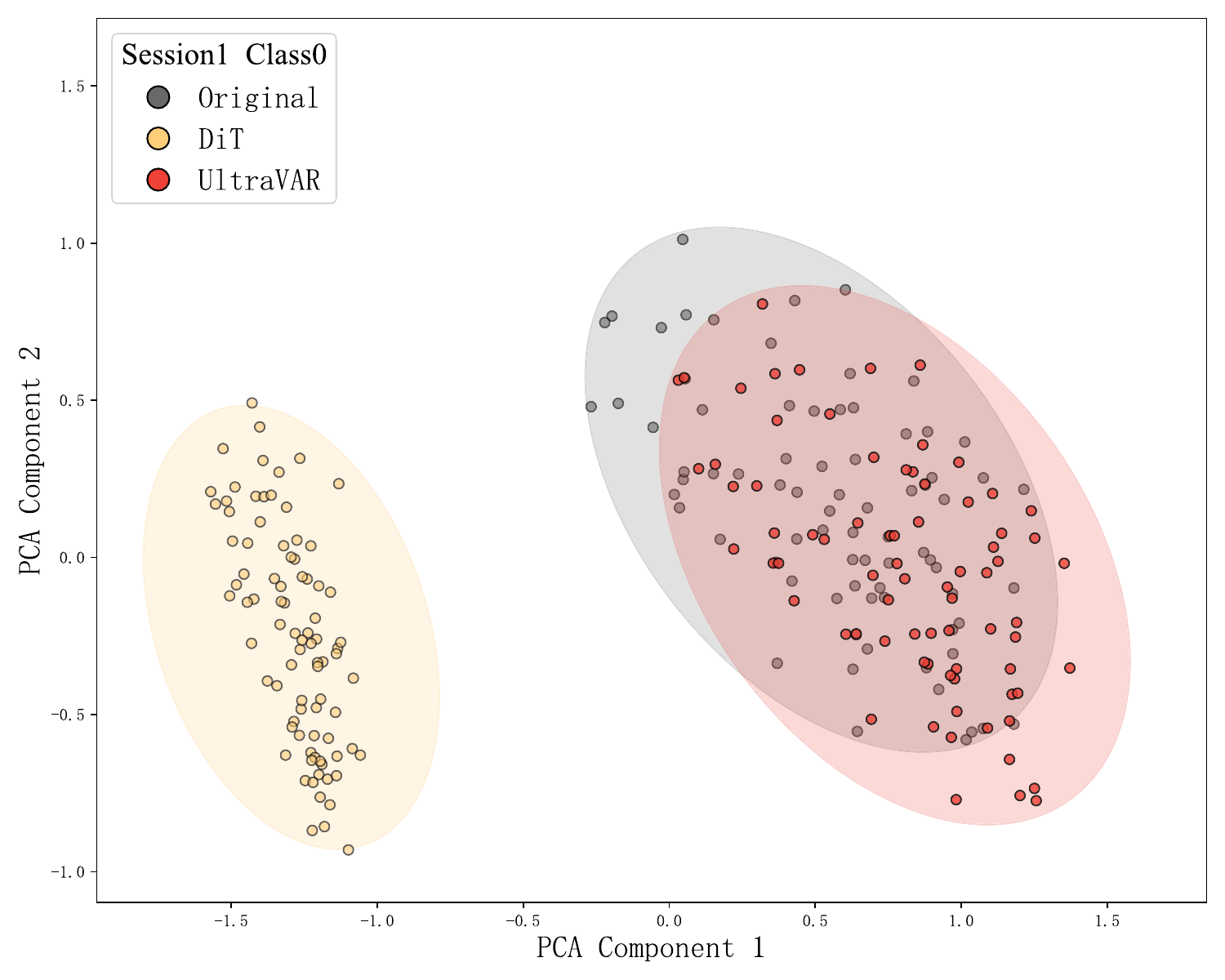}
    \includegraphics[width=0.48\linewidth]{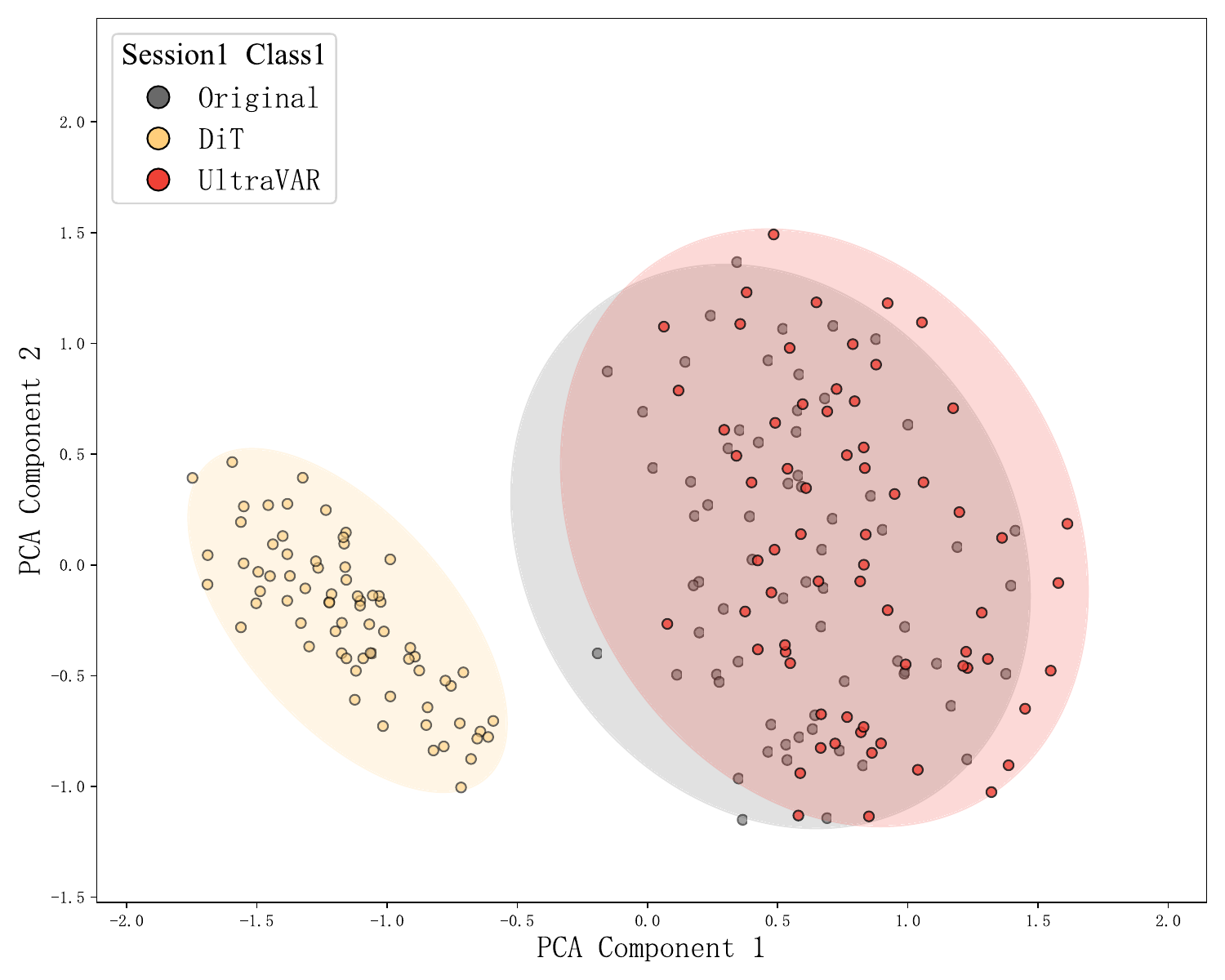}
    \includegraphics[width=0.48\linewidth]{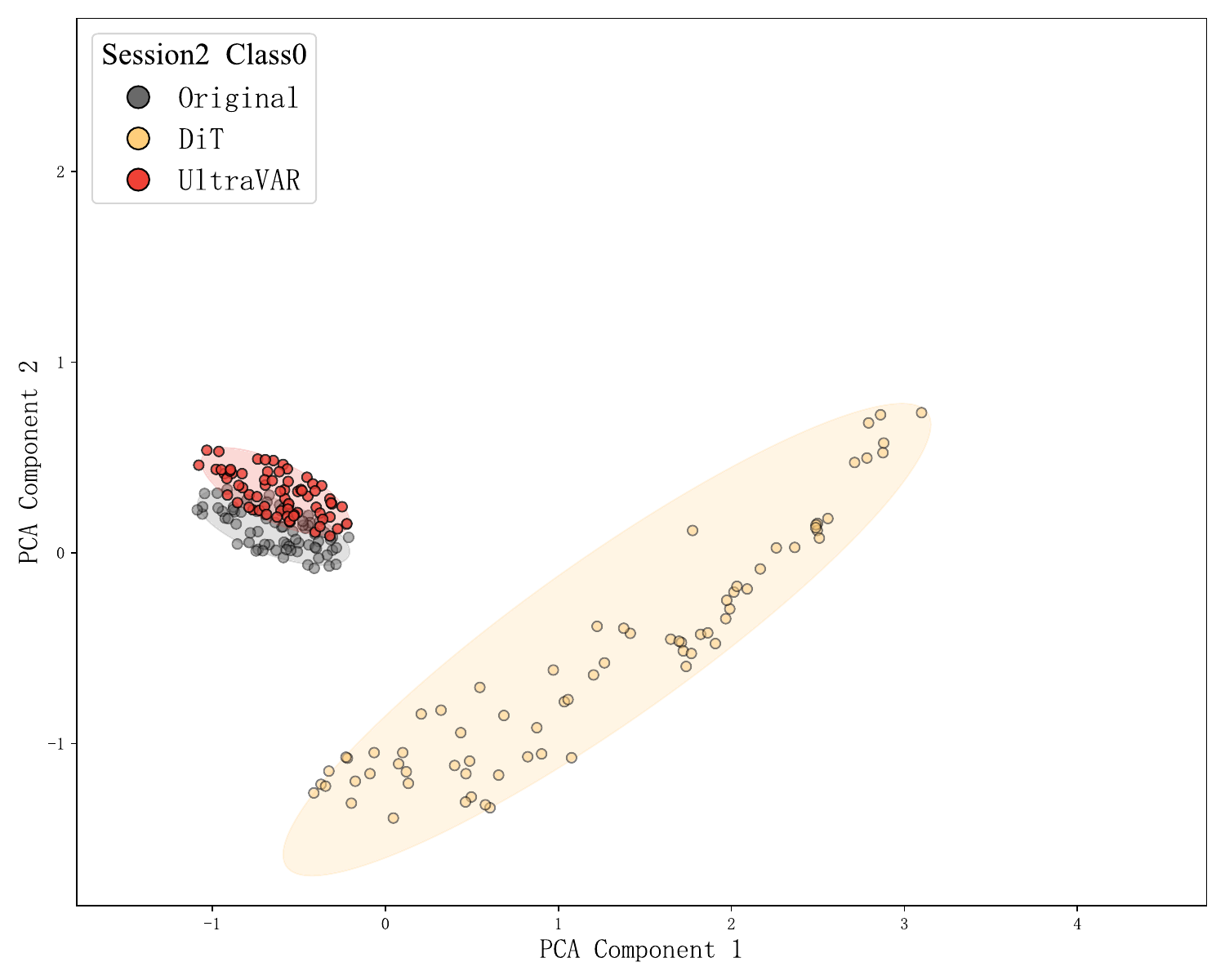}
    \includegraphics[width=0.48\linewidth]{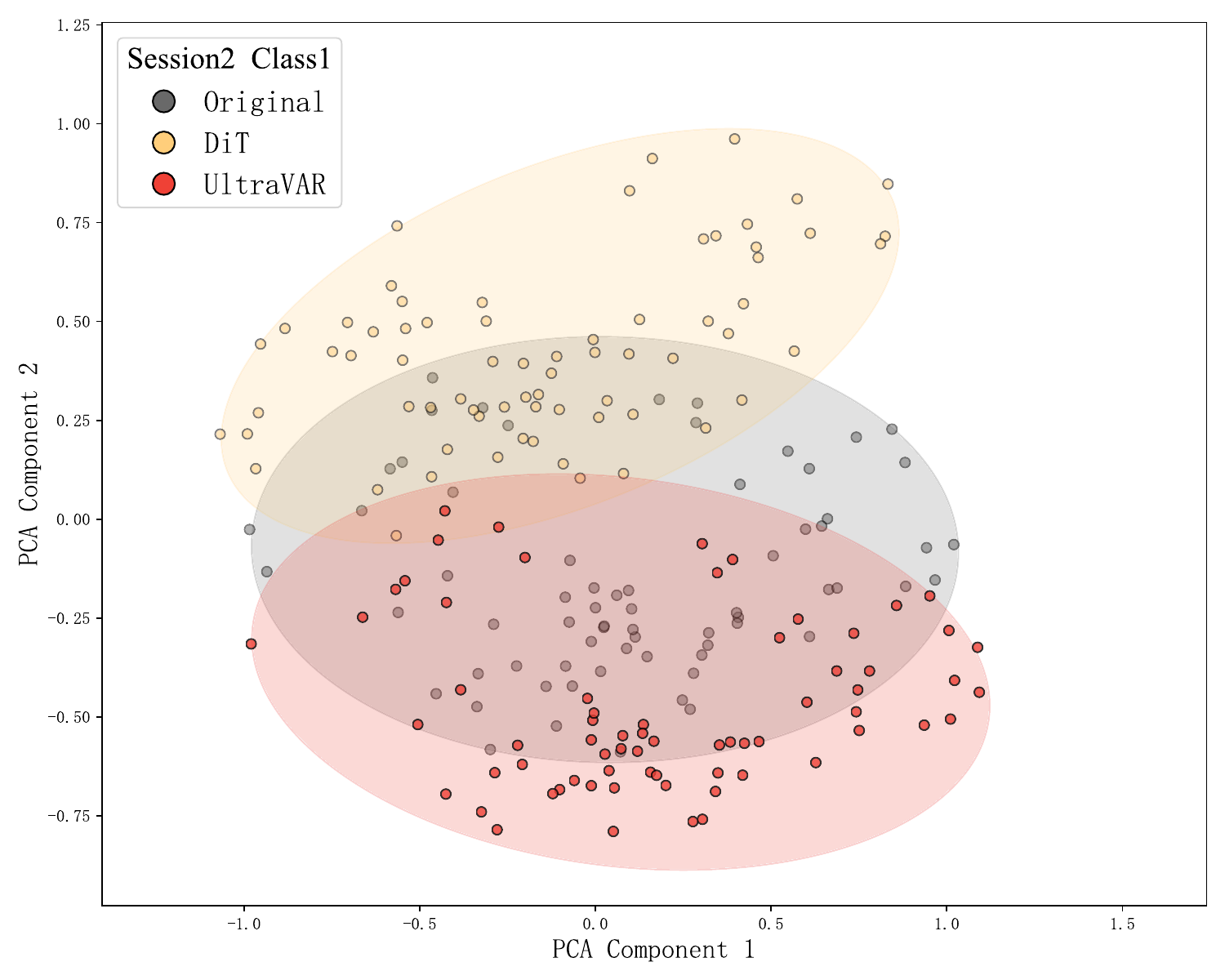}
    \caption{PCA visualization comparing feature distributions of original fUS data, DiT-generated data, and UltraVAR-generated data. The closer clustering of UltraVAR samples to the original data, relative to DiT samples, indicates that UltraVAR generates data with representations more faithful to the original dataset.}
    \label{fig:tsne}
\end{figure*}

\subsection{Dataset and Preprocessing}
For experimental validation, we utilized the publicly available fUS imaging dataset of the human brain introduced by Rabut \etal \cite{rabut2024functional}. This dataset is particularly suitable for our study due to its clinical relevance, involving human subjects and tasks designed to elicit distinct brain activity patterns, which is crucial for evaluating the ability of our generative model to capture physiologically meaningful variations. The data acquisition protocol comprised two distinct experimental sessions: guitar playing (Session1) and a line-connecting game (Session2). Each session employed a block design wherein participants alternated between 100-second rest phases (designated Class0) and 50-second task-active phases (designated Class1). This block design is instrumental as it provides clearly delineated periods of differing neural states, essential for training and evaluating a classifier on the generated and original fUS data.

To rigorously assess the generalizability and robustness of our proposed method, one complete task phase from each session was deliberately held out as an independent test set. This strategy is more stringent than random sampling for test data, as it ensures that the model is evaluated on a continuous block of data it has not encountered during training, better reflecting real-world scenarios where a model might be applied to new, unseen experimental runs. The data from the remaining phases constituted the training and validation sets.

Preprocessing involved temporal slicing of the continuous fUS data streams. This approach converts the time-series fUS data into individual image frames, each representing a snapshot of cerebrovascular activity. This resulted in 720 training images and 270 test images for Session1, and 1530 training images and 270 test images for Session2. The difference in training set sizes between Session1 and Session2 reflects the varying number of task/rest blocks available after holding out the test phase. All extracted image frames were subsequently cropped to a uniform resolution of $128 \times 128$ pixels. This resolution was chosen as a balance between retaining sufficient spatial detail of the vascular networks and computational feasibility for training complex generative models. Cropping ensures uniformity in input size, a standard requirement for most deep learning architectures, and can also help focus the model on relevant brain regions if the original field of view contains extraneous areas.

\begin{table*}[ht]
\caption{Quantitative evaluation of downstream classification performance. This table presents the Accuracy, Precision, Recall, and F1-Score for Session1 and Session2 test sets, comparing results obtained using the original data against data augmented by various generative models. The $\uparrow$ symbol indicates that higher values represent better performance. The proposed method demonstrates superior results across most metrics and both sessions.}
\centering
\begin{tabular}{c|cccc|cccc}
\toprule
\multirow{2}{*}{Session} & \multicolumn{4}{c|}{Session1}               & \multicolumn{4}{c}{Session2}                \\ \cmidrule(lr){2-9} 
                         & Accuracy $\uparrow$& Precision $\uparrow$& Recall $\uparrow$& F1-Score $\uparrow$   & Accuracy $\uparrow$& Precision $\uparrow$& Recall $\uparrow$& F1-Score $\uparrow$   \\ \midrule
Original                 & 0.800      & 0.800       & 0.533  & 0.64  & 0.755    & 0.667     & 0.533  & 0.593 \\
CVAE~\cite{Sohn2015LearningSO}                      & 0.867    & 0.846     & 0.733  & 0.786 & 0.8      & 0.750      & 0.600    & 0.667 \\
VQVAE~\cite{Oord2017NeuralDR}                    & 0.778    & 0.727     & 0.533  & 0.617 & 0.733    & 0.615     & 0.533  & 0.571 \\
BIGGAN~\cite{brock2018large}                   & 0.809    & 0.835     & 0.533  & 0.651 & 0.755    & 0.833     & 0.333  & 0.476 \\
StyleGAN-XL~\cite{sauer2022stylegan}               & 0.837    & 0.809     & 0.667  & 0.731 & 0.778    & 0.778     & 0.467  & 0.583 \\
VQGAN~\cite{esser2021taming}                    & 0.800      & 0.800       & 0.533  & 0.64  & 0.689    & 0.529     & 0.533  & 0.581 \\
DiT~\cite{peebles2023scalable}                      & 0.767    & 0.667     & 0.600    & 0.632 & 0.725    & 0.615     & 0.467  & 0.438 \\
\textbf{Proposed}                 & \textbf{0.889}    & \textbf{0.857}     & \textbf{0.800}    & \textbf{0.828} & \textbf{0.822}   & \textbf{0.889}     & \textbf{0.600}    & \textbf{0.667} \\ \bottomrule
\end{tabular}
\label{tab:clf}
\end{table*}
\subsection{Implementation Details}
The proposed UltraVAR model was implemented utilizing the PyTorch framework, a widely adopted open-source machine learning library known for its flexibility and robust support for GPU acceleration, which is critical for training deep generative models. The model was trained on a system equipped with 2 NVIDIA A800 GPUs, providing substantial computational power necessary for handling the complexity of the visual auto-regressive model and the dataset size. Input images were resized to $128 \times 128$ pixels, consistent with the preprocessing step.

Training proceeded for 200 epochs with a batch size of 4. The number of epochs was chosen to ensure sufficient convergence of the model, while the batch size was determined by GPU memory constraints and the desire to have stable gradient estimates. We employed the AdamW optimizer \cite{loshchilov2017fixing}, an extension of the Adam optimizer that improves weight decay regularization, with hyperparameters set to $\beta_1 = 0.9$ and $\beta_2 = 0.999$. These are standard, well-tested values for Adam-like optimizers. This was coupled with a cosine annealing learning rate scheduler \cite{loshchilov2016sgdr}. This scheduler starts with a relatively higher learning rate which is gradually decreased following a cosine function, helping the model to escape local minima early in training and then fine-tune its weights more precisely as it approaches convergence, thus promoting training stability and often leading to better final performance.

During the inference phase, images were processed at the same $128 \times 128$ resolution, ensuring consistency between the training and evaluation pipelines. This is crucial because discrepancies in resolution or preprocessing between training and inference can lead to performance degradation. Crucially, no conventional data augmentation techniques (\eg, geometric transformations like rotation or flipping, noise injection) were applied during the training of UltraVAR itself. This decision was made to preserve the intrinsic characteristics of the original data distribution. By avoiding these common augmentation methods, we can facilitate an unbiased assessment of the generative model's intrinsic capability to synthesize realistic and diverse fUS data, rather than evaluating its ability to learn from artificially expanded, and potentially distorted, versions of the original samples. The goal is for UltraVAR to learn the underlying data manifold purely from the original samples and then generate novel samples from that learned manifold.

\subsection{Comparison With Other Existing Synthetic Methods}
To evaluate the efficacy of the proposed UltraVAR framework for high-fidelity fUS image reconstruction, we performed a comprehensive comparative analysis against several representative state-of-the-art generative models. The baseline methods included: Conditional Variational Autoencoder (CVAE) \cite{Sohn2015LearningSO}, Vector Quantized Variational Autoencoder (VQVAE) \cite{Oord2017NeuralDR}, BigGAN \cite{brock2018large}, StyleGAN-XL \cite{sauer2022stylegan}, Vector Quantized Generative Adversarial Network (VQGAN) \cite{esser2021taming}, and Diffusion Transformer (DiT) \cite{peebles2023scalable}. These models represent a diverse range of generative approaches, including VAEs, GANs, and diffusion models, providing a robust benchmark for UltraVAR.

A qualitative visual assessment, presented in \cref{fig:vis}, underscores the superiority of UltraVAR. The synthetic images generated by UltraVAR exhibit remarkable visual fidelity, closely mirroring the original fUS data in terms of texture, contrast, and the intricate structure of the underlying vascular networks. Specifically, UltraVAR maintains the continuity and branching patterns of microvessels, which are critical for interpreting fUS images. Notably, UltraVAR demonstrates a strong capability to preserve the subtle neurovascular coupling features that differentiate between brain activity states (Class0 vs. Class1) across both experimental sessions. These features might include localized changes in blood volume or flow patterns that are indicative of neural activity. In contrast, images generated by the baseline methods frequently exhibit noticeable artifacts such as unnatural smoothness, pixelation, or structural inconsistencies. For instance, samples from VQVAE and VQGAN occasionally lack sharpness or exhibit blocky artifacts due to the quantization step, whereas GAN-based approaches like BigGAN and StyleGAN-XL sometimes introduce unrealistic structural patterns or overly smoothed textures inconsistent with the complex, somewhat noisy appearance of genuine brain vasculature. DiT, while powerful, might also struggle with fine-grained details specific to fUS if not perfectly tuned. The ability of UltraVAR to avoid these pitfalls is likely due to its auto-regressive nature combined with the Smooth Scaling Layer and Perception Enhancement Module, which are designed to maintain coherence and perceptual quality.

Quantitative evaluations, depicted in \cref{fig:comp}, further corroborate these visual findings. We employed established image quality metrics: the Structural Similarity Index Measure (SSIM) \cite{wang2004image}, the Multi-Scale Structural Similarity Index Measure (MS-SSIM) \cite{wang2003multiscale}, and the Fr\'{e}chet Inception Distance (FID) \cite{heusel2017gans}. SSIM measures similarity based on luminance, contrast, and structure, while MS-SSIM extends this by evaluating these aspects at multiple scales, making it more robust to variations in viewing conditions and image resolution. FID assesses the similarity between the distributions of generated images and real images in a feature space derived from a pre-trained Inception network; lower FID scores indicate a closer match between the two distributions, suggesting higher perceptual quality and diversity. Across both Session1 and Session2 datasets, UltraVAR consistently achieved the highest SSIM and MS-SSIM scores and, conversely, the lowest FID values among all evaluated models. Specifically, for Session1, UltraVAR yielded an SSIM of 0.882, an MS-SSIM of 0.971, and an FID of 70.496. For Session2, the corresponding metrics were 0.978 (SSIM), 0.995 (MS-SSIM), and 47.194 (FID). These quantitative results affirm that UltraVAR generates images possessing superior structural integrity (high SSIM/MS-SSIM) and perceptual quality that more faithfully captures the characteristics of the original fUS data distribution (low FID) compared to competing state-of-the-art methods. The significant margins in these metrics suggest that UltraVAR's architectural innovations directly contribute to more realistic image reconstruction.

Furthermore, we investigated the fidelity of the learned feature representations using Principal Component Analysis (PCA) visualization, as illustrated in \cref{fig:tsne}. PCA is a dimensionality reduction technique that projects high-dimensional data (in this case, image features) onto a lower-dimensional space while preserving as much variance as possible. This allows for visualization of the relationships between different data distributions. This analysis compares the feature distributions extracted from the original fUS images against those generated by UltraVAR and a strong baseline, DiT. The visualization clearly reveals that the feature cluster corresponding to UltraVAR-generated samples exhibits significantly greater overlap and proximity to the cluster representing the original data features, relative to the DiT-generated samples. This observation indicates that UltraVAR not only synthesizes visually convincing images but also captures the underlying semantic feature distribution of the authentic fUS data more faithfully. A closer match in feature space implies that UltraVAR is learning the complex, high-dimensional manifold of real fUS data more accurately, which is crucial for generating samples that are not just superficially similar but also share deeper statistical properties with the original data. This fidelity in feature space is a strong indicator of the synthetic data's potential utility in downstream tasks, as models trained on such data are more likely to learn relevant patterns. Collectively, these visual, quantitative, and feature-space analyses underscore UltraVAR's advanced capability in generating diverse yet realistic fUS images that effectively preserve critical physiological information.
\begin{figure*}[ht]
    \centering
    \includegraphics[width=\linewidth]{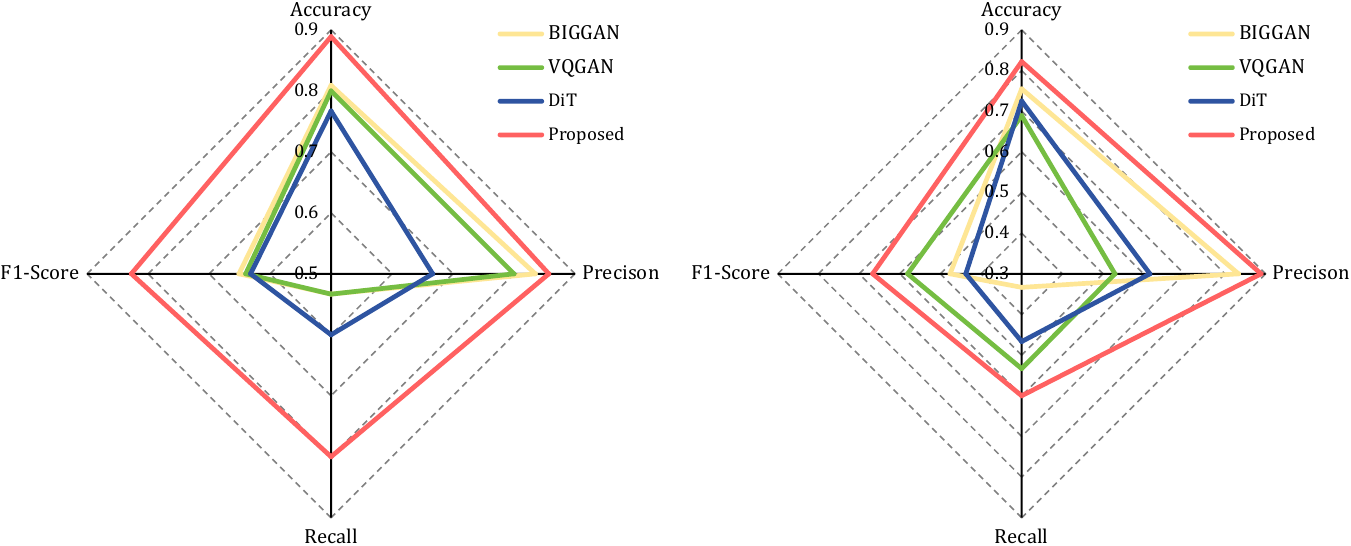}
    \caption{Performance comparison on the downstream classification task. The radar charts illustrate classification metrics for different generative models on Session1 (left) and Session2 (right), with the proposed UltraVAR model showing superior performance.}
    \label{fig:radar}
\end{figure*}

\begin{figure*}[ht]
    \centering
    \includegraphics[width=\linewidth]{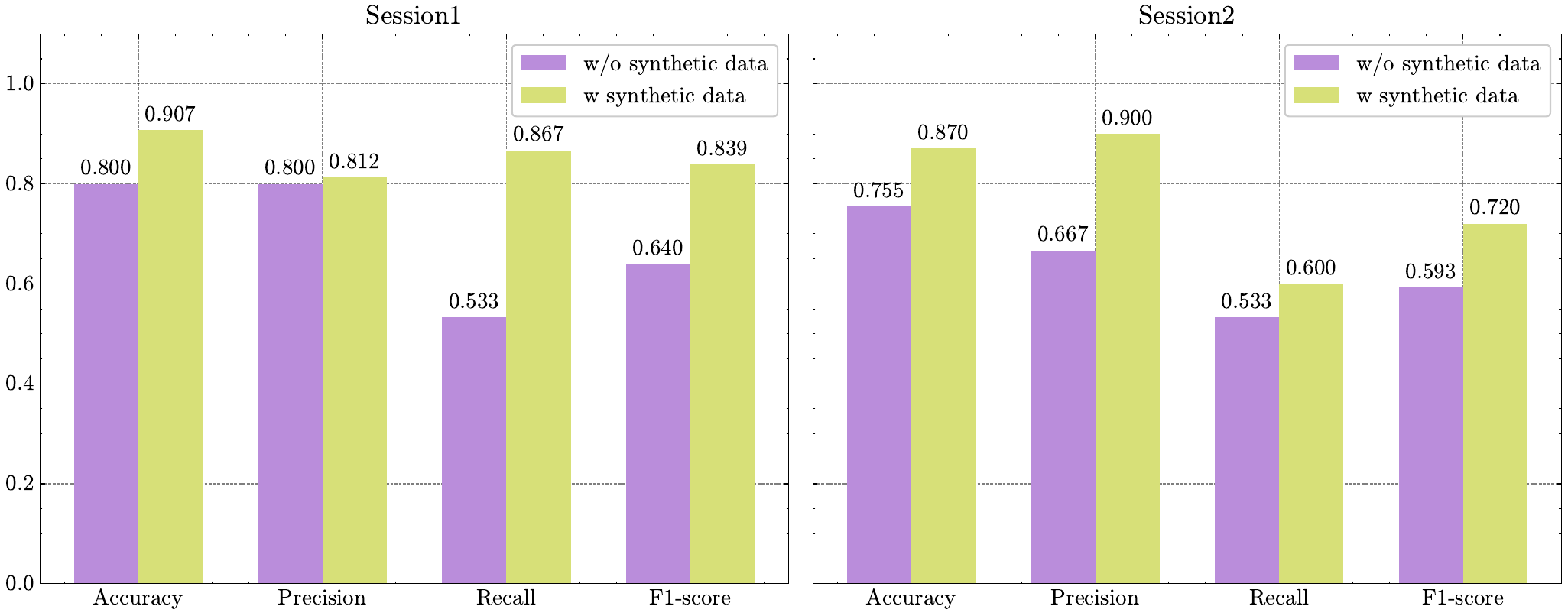}
    \caption{Enhancement of downstream classification performance through UltraVAR data augmentation, indicating improved model fairness. The bar charts compare Accuracy, Precision, Recall, and F1-Score for models trained solely on the original limited dataset (w/o synthetic data) versus models trained with additional synthetic data generated by UltraVAR (w synthetic data). Results are shown for both Session1 (left) and Session2 (right).}
    \label{fig:fair}
\end{figure*}

\begin{figure*}[ht]
    \centering
    \includegraphics[width=\linewidth]{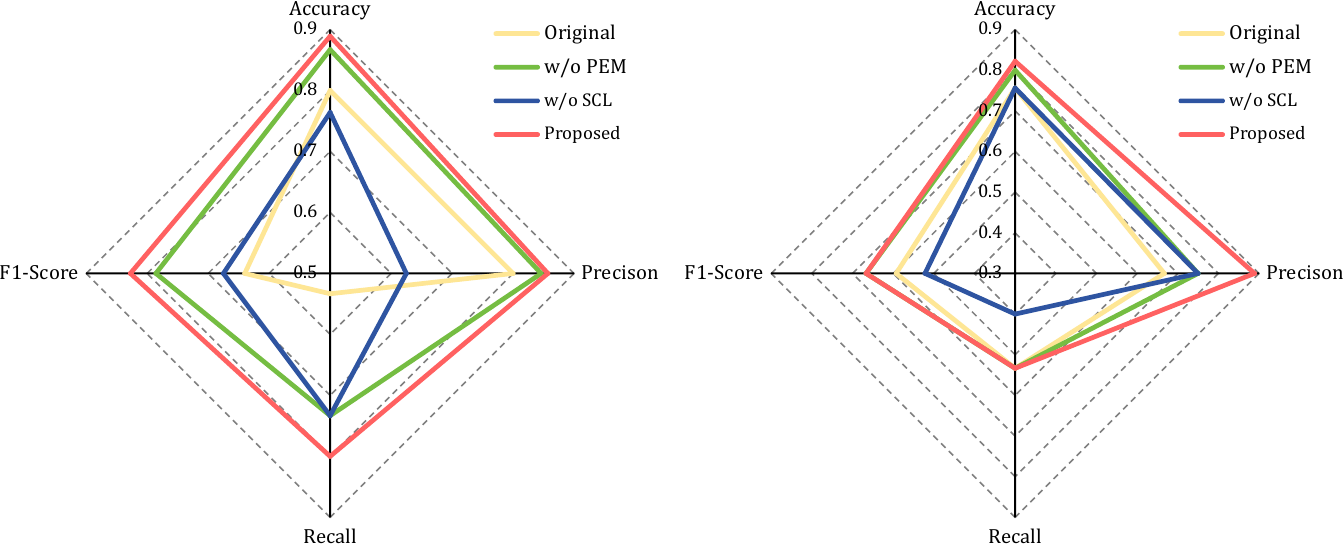}
    \caption{Ablation study results comparing classification performance. The radar charts show the impact of removing key components (PEM or SCL) from the proposed UltraVAR framework on Session1 (left) and Session2 (right), demonstrating the contribution of each module to overall performance.}
    \label{fig:ab}
\end{figure*}

\begin{table*}[ht]
\caption{Quantitative evaluation of ablation study. This table presents classification metrics for the original data and three variants of the proposed method: without Perception Enhancement Module (w/o PEM), without Smooth Scaling Layer (w/o SCL), and the complete proposed UltraVAR framework, demonstrating the contribution of each component to overall performance across both sessions.}
\centering
\begin{tabular}{c|cccc|cccc}
\toprule
\multirow{2}{*}{Session} & \multicolumn{4}{c|}{Session1}               & \multicolumn{4}{c}{Session2}                \\ \cmidrule(lr){2-9} 
                         & Accuracy $\uparrow$& Precision $\uparrow$& Recall $\uparrow$& F1-Score $\uparrow$   & Accuracy $\uparrow$& Precision $\uparrow$& Recall $\uparrow$& F1-Score $\uparrow$   \\ \midrule
Original                 & 0.800      & 0.800       & 0.533  & 0.640  & 0.755    & 0.667     & 0.533  & 0.593 \\
w/o PEM & 0.867 & 0.846 & 0.733 & 0.786 & 0.800 & 0.750 & 0.600 & 0.667 \\
w/o SCL & 0.764 & 0.625 & 0.733 & 0.675 & 0.756 & 0.750 & 0.400 & 0.522 \\
\textbf{Proposed}                 & 0.889    & 0.857     & 0.800    & 0.828 & 0.822   & 0.889     & 0.600    & 0.667 \\ \bottomrule
\end{tabular}
\label{tab:ab}
\end{table*}

\subsection{Downstream Task Evaluation}
A critical measure of the utility of synthetic data lies in its ability to enhance the performance of downstream tasks. We evaluated the effectiveness of UltraVAR-generated data by using it for a classification task aimed at distinguishing between rest (Class0) and task-active (Class1) brain states based on fUS images. The performance of a standard classifier trained on datasets augmented by UltraVAR and other baseline generative models was compared against its performance when trained solely on the original data. The chosen metrics—Accuracy, Precision, Recall, and F1-Score—provide a comprehensive view of classifier performance, especially important when class imbalances might exist or when false positives and false negatives have different implications.

The results, summarized in \cref{tab:clf}and visualized in \cref{fig:radar}, demonstrate the significant benefit conferred by UltraVAR augmentation. Across both Session1 and Session2, the classifier trained with UltraVAR-augmented data achieved superior performance compared to models trained on the original data or data augmented by other methods. Specifically, for Session1, UltraVAR augmentation boosted accuracy from 0.800 (original) to 0.889, Precision from 0.800 to 0.857, Recall from 0.533 to 0.800, and F1-Score from 0.640 to 0.828. Similar substantial improvements were observed for Session2, with accuracy increasing from 0.755 to 0.822 and F1-Score from 0.593 to 0.667. The notable increase in Recall for Session1 (from 0.533 to 0.800) is particularly important, as it indicates that the model becomes much better at identifying the task-active state, which is often the class of interest. This suggests that UltraVAR generates synthetic samples that effectively represent the characteristics of the task-active class, helping the classifier to learn more robust decision boundaries. Notably, UltraVAR consistently outperformed other augmentation strategies like CVAE, VQVAE, BigGAN, StyleGAN-XL, VQGAN, and DiT across most metrics, highlighting the higher quality and relevance of the synthetic samples it generates. This superior performance is likely attributable to UltraVAR's enhanced ability to preserve neurovascular coupling features, which are the key differentiators between the brain states.

Beyond performance enhancement, the utilization of UltraVAR addresses the challenges of fairness and robustness in model evaluation, particularly pertinent given the inherent data scarcity in fUS imaging stemming from ethical and procedural constraints. Original fUS datasets are frequently limited in size and may lack diversity, creating a potential for biased models that exhibit poor generalization capabilities. A model trained on such limited data might perform well on subjects or conditions similar to the training set but fail significantly on others.

To systematically evaluate the impact of UltraVAR on these aspects, we utilized data acquired from Session1 and Session2. We assigned each session with 270 fUS images, categorized into rest phases (Class0: 180 images) and task-active phases (Class1: 90 images). This setup reflects a common scenario of class imbalance, where one state is more prevalent or easier to acquire data for than another. Adhering to standard data partitioning practices, we randomly allocated 80\% of the data for training (216 images: 141 Class0, 75 Class1) and reserved the remaining 20\% for testing (54 images: 39 Class0, 15 Class1).

As illustrated in \cref{fig:fair}, augmenting the limited original training data with synthetic samples generated by UltraVAR yields marked improvements across all standard classification metrics, including Accuracy, Precision, Recall, and F1-Score. UltraVAR facilitates this enhancement by generating diverse yet physiologically plausible fUS images. This diversity could encompass subtle variations in signal intensity, minor shifts in vascular patterns that are still within physiological norms, or variations in the expression of neurovascular coupling features. These variations effectively expand the training distribution while meticulously preserving essential neurovascular coupling features. By exposing the classifier to a broader spectrum of realistic data variations during training, UltraVAR augmentation mitigates the risk of overfitting to the constrained original samples and demonstrably enhances the model's ability to generalize to unseen data.

Notably, this process directly contributes to improving model fairness, particularly in the context of imbalanced classes. Recognizing the inherent class imbalance within the original data (Class1 being the minority), we augmented the training set with 66 additional synthetic Class1 samples, ensuring these samples maintained physiological plausibility. This targeted augmentation, designed to counterbalance the empirical data distribution by specifically increasing the representation of the minority class, aligns with established practices for evaluating and mitigating selection rate disparities in fairness assessments. Consequently, UltraVAR not only reduces the risk of biases arising from data scarcity but also promotes greater equity in prediction patterns across different task paradigms. This leads to the development of more robust and reliable fUS-based classification systems suitable for demanding clinical applications such as BCIs \cite{tang2023flexible,gao2021interface,9310231} and neuromodulation \cite{lee2012neuromodulation,knotkova2021neuromodulation,ryvlin2021neuromodulation}, where consistent performance across various conditions and patient populations is paramount.

\subsection{Ablation Studies}
To investigate the individual contributions of the novel components integrated within UltraVAR—specifically, the Smooth Scaling Layer (SCL) and the Perception Enhancement Module (PEM)—we conducted a systematic ablation study. This involves removing each component individually and observing the impact on performance, thereby isolating its effect. We evaluated the downstream classification performance of three model variants: (1) the full UltraVAR framework, (2) UltraVAR without the PEM (denoted `w/o PEM'), and (3) UltraVAR without the SCL (denoted `w/o SCL'). Performance was benchmarked against using only the original data.

The quantitative outcomes of this ablation study are detailed in \cref{tab:ab}, with a comparative visualization provided in the radar charts of \cref{fig:ab}. The findings clearly demonstrate the distinct importance of both the SCL and PEM modules.

Removal of the PEM (w/o PEM) resulted in a discernible decrease in performance compared to the full model across both sessions, although this variant still outperformed the baseline using only original data. For instance, in Session1, the F1-score dropped from 0.828 (Proposed) to 0.786 (w/o PEM). This suggests that the PEM, by refining features such as contrast, sharpness, and overall visual fidelity, effectively enhances the perceptual quality of the generated images in a way that makes critical physiological features more discernible for the downstream classifier. While the VQVAE decoder reconstructs the main structures, PEM likely fine-tunes these reconstructions to suppress minor artifacts or enhance subtle details that are beneficial for classification.

The removal of the SCL (w/o SCL) led to a more substantial degradation in classification performance. In Session1, the F1-score decreased significantly to 0.675, while in Session2, it dropped markedly to 0.522. This highlights the critical role of the SCL in maintaining information integrity and spatial coherence throughout the multi-scale, hierarchical reconstruction process. The SCL ensures smooth transitions and consistency between image patches generated at different resolutions. Without it, the generated images might suffer from discontinuities or misalignments between regions, disrupting the representation of continuous vascular networks. This loss of structural integrity would directly impact the model's ability to learn and represent neurovascular dynamics accurately, leading to poorer classification of brain states. Preserving these vital spatial topological relationships of vascular networks is essential for accurately representing neurovascular dynamics and enabling effective subsequent classification.

Comparing the variants, the complete UltraVAR framework consistently yielded the optimal performance across all evaluated metrics in both sessions. This confirms that the Smooth Scaling Layer and the Perception Enhancement Module provide synergistic benefits. The SCL ensures the foundational structural accuracy and coherence of the generated fUS images, particularly in maintaining the integrity of vascular networks across scales. The PEM then builds upon this by refining the perceptual quality, making the anatomically accurate structures more visually clear and suppressing any residual artifacts. Each module contributes significantly to UltraVAR's capacity to generate high-quality, physiologically relevant fUS images. These generated images effectively augment limited training data, leading to improved downstream task performance. The ablation study thus validates our design choices and underscores the efficacy of the proposed architectural enhancements within the UltraVAR framework.

\section{Discussion}
\subsection{Comparison with State-of-the-Art Methods}
Our experimental results demonstrate that UltraVAR significantly outperforms existing generative approaches for functional ultrasound image reconstruction and augmentation. While other generative models have been successfully applied to various imaging domains \cite{Sohn2015LearningSO,Oord2017NeuralDR,brock2018large,sauer2022stylegan,esser2021taming,peebles2023scalable}, they exhibit limitations when applied specifically to fUS imaging. These limitations primarily stem from their inability to preserve the intricate neurovascular coupling features and spatial topological relationships that are essential for accurate fUS interpretation.

The visual assessment in \cref{fig:vis} reveals that existing methods frequently reconstruct images with noticeable artifacts such as unnatural smoothness (seen in StyleGAN-XL outputs), pixelation (evident in VQVAE and VQGAN samples), or structural inconsistencies that disrupt the continuous nature of vascular networks. In contrast, UltraVAR successfully maintains the complex branching patterns and microvascular structures that characterize authentic fUS images. The quantitative metrics in \cref{fig:comp} further validate these observations, with UltraVAR achieving superior performance across SSIM, MS-SSIM, and FID scores for both experimental sessions.

The feature space analysis through PCA visualization (\cref{fig:tsne}) provides additional evidence that UltraVAR captures the underlying data distribution more faithfully than competing methods. The closer alignment between UltraVAR-reconstructed samples and original data in the feature space indicates that our approach preserves the semantic characteristics of fUS images better than alternative methods. This preservation of semantic features is crucial for downstream applications that rely on accurate representation of neurovascular dynamics, particularly in cross-subject scenarios where model generalization is essential \cite{dong2024multi}.

\subsection{Clinical Relevance for Fetal Ultrasound and Future Directions}
The clinical implications of UltraVAR are substantial, particularly when considering its potential adaptation to fetal ultrasound imaging. Data scarcity and the need for expert interpretation are significant hurdles. UltraVAR's proven ability to augment fUS datasets and improve downstream classification accuracy (\cref{tab:clf} and \cref{fig:radar}) suggests its potential to develop intelligent system for fetal diagnostics. By constructing diverse yet physiologically plausible fUS images, UltraVAR can help create intelligent systems that are less prone to biases from limited datasets and can generalize better, potentially leading to earlier and more reliable detection of fetal abnormalities \cite{wang2025smart}.

The enhancement of model fairness, as demonstrated in our experiments (\cref{fig:fair}) by addressing class imbalance and improving performance on limited data, is also clinically significant. This ensures that intelligent systems developed for fetal imaging can perform reliably across diverse populations and varied presentations of conditions. The high-quality synthetic data reconstructed by UltraVAR could serve as a robust foundation for developing advanced intelligent systems, including those for automatic segmentation of fetal organs, predictive analytics for fetal growth and development, and improved image reconstruction.

Future directions should focus on adapting and validating UltraVAR for direct application in fetal ultrasound. This includes: (1) Investigating the potential for UltraVAR to contribute to real-time processing and analysis systems for fetal ultrasound. (2) Extending the UltraVAR framework to handle 3D/4D fetal ultrasound data, which provides more comprehensive anatomical information.

\subsection{Limitations}
Despite its promising results, UltraVAR has limitations that warrant discussion. 
While UltraVAR addresses data scarcity issues, the quality of reconstructed samples remains dependent on the diversity of the original training dataset. The limited availability of fUS data from diverse populations, particularly in clinical contexts, may constrain the model's ability to generate samples that fully represent the variability in different demographic groups, potentially introducing biases in downstream applications. 
The hierarchical nature of UltraVAR's reconstruction process, coupled with the Smooth Scaling Layer and Perception Enhancement Module, increases computational complexity compared to simpler generative models. This may limit real-time applications or deployment on resource-constrained devices, which is particularly relevant for point-of-care ultrasound applications.

\section{Conclusion}
This paper presents UltraVAR, the first generative AI framework for functional ultrasound image augmentation that addresses critical challenges in fUS imaging. By leveraging a pre-trained visual auto-regressive model enhanced with the proposed Smooth Scaling Layer and Perception Enhancement Module, UltraVAR generates diverse, physiologically plausible ultrasound images that preserve essential neurovascular coupling features and spatial topological relationships of vascular networks. The experimental validation demonstrates that datasets augmented with UltraVAR yield significant improvements in downstream classification accuracy compared to conventional augmentation methods, establishing a high-quality foundation for ultrasound-based neuromodulation techniques and brain-computer interface technologies. The framework's ability to maintain critical physiological correlations while enhancing model fairness represents a substantial advancement for functional ultrasound imaging, potentially accelerating its clinical translation and expanding its applications in neurophysiological research, from neonatal brain development monitoring to brain function decoding.

\bibliographystyle{IEEEtranDOI}
\bibliography{reference}

\end{document}